# Superionic phase transition of copper(I) sulfide and its implication for purported superconductivity of LK-99

*Prashant K. Jain\*, a, b*

aDepartment of Chemistry, University of Illinois Urbana-Champaign, Urbana, Illinois 61801, United States

bMaterials Research Laboratory, University of Illinois Urbana-Champaign, Urbana, Illinois 61801, United States

\*Corresponding author: Prof. Prashant K. Jain

E-mail: jain@illinois.edu.

Lee, Kim, and co-workers have reported in a publication[1] and two preprints[2,3] a modified lead apatite material, named LK-99 after Lee and Kim, with the putative structure, $Pb_9Cu(PO_4)_6O$, and purported ambient temperature and pressure superconductivity. If this claim were to be validated, it would qualify as a historic breakthrough in condensed matter physics and all of physical science and technology. After all, superconductivity is a phenomenon inherent to low temperatures or high pressures. Achieving resistance-less electrical conductivity at room temperature and ambient pressure could in principle pave the way for practical, economically viable applications of superconductors in higher-efficiency electrical transmission lines, quantum computing hardware that does not require cryogenic cooling, high-speed data interconnects, computer microchips with orders-of-magnitude lower power consumption, and powerful electromagnets for magnetic-levitation trains and magnetic resonance imaging. Most pivotally, the discovery of such a material would open the floodgates for a new class of high-temperature superconductors and instigate the search for new models of electron pairing underlying superconductivity.

Although there remains uncertainty about the structure of LK-99, X-ray diffraction (XRD) unambiguously shows the presence of copper(I) sulfide byproduct in the modified lead apatite. The formation of copper(I) sulfide as a byproduct has been verified by a replication of the synthesis by Awana and co-workers.[4]

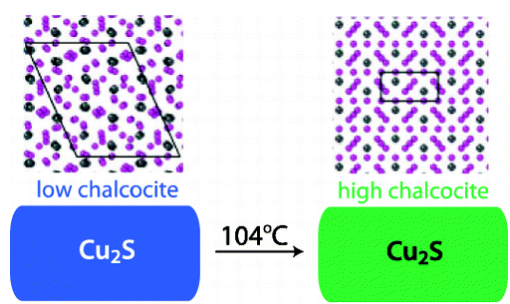

**Figure 1**. Known solid–solid phase transition of bulk $Cu_2S$ at 104 °C as depicted in a figure adapted from a previous publication by the author and his co-workers. Reprinted with permission from ref 8. Copyright 2011 American Chemical Society.

There are some properties of copper(I) sulfide, which are relevant to the current discussion. First, Cu is in the 1+ oxidation state with a $d^{10}$ configuration, making copper(I) sulfide a diamagnetic solid.[5] Secondly, bulk $Cu_2S$ has a known solid–solid phase transition at 104 °C (Figure 1): the low-temperature γ phase (low chalcocite), with a monoclinic/pseudo-orthorhombic crystal structure[6] undergoes a transformation to the high-temperature β phase (high chalcocite), with a hexagonal crystal structure.[7–9] This γ-to-β phase transition temperature is nearly coincident with the temperature of 104.8 °C at which LK-99 is claimed[1] to exhibit an abrupt order-of-magnitude jump in resistivity (Figure 2, left).[1]

Given this coincidence, it is necessary to examine the phase transition of $Cu_2S$ even further. The β phase is superionic and has a disordered arrangement of $Cu^+$ cations within a crystalline $S^{2-}$ anionic framework, whereas in the γ phase, there are well-defined lattice sites for Cu. This structural change, effectively a melting of the Cu sublattice, is manifested in a sharp rise in the ionic conductivity of bulk $Cu_2S$ at 104 °C.[10,11] What about the electronic contribution to the conductivity? Although in high purity stoichiometric $Cu_2S$, electronic conductivity is low, this is not the case for samples that have been exposed to air. Air exposure, even to a mild degree, induces the formation of copper oxide leading to a Cu-deficient $Cu_{2-\delta}S$ phase (djurleite).[6,7,12,13] Ionized Cu vacancies in the lattice give rise to free holes in the valence band of $Cu_{2-\delta}S$. Therefore, $Cu_{2-\delta}S$ is p-type with free hole concentrations as high as $10^{21}$ $cm^{-3}$—established in studies on nanoparticles by the emergence of a localized surface plasmon resonance[7,12,13]—and a substantial electronic conductivity. The electronic conductivity dwarfs the ionic conductivity[9] and is therefore the dominant contributor to the total (electronic + ionic) electrical conductivity.

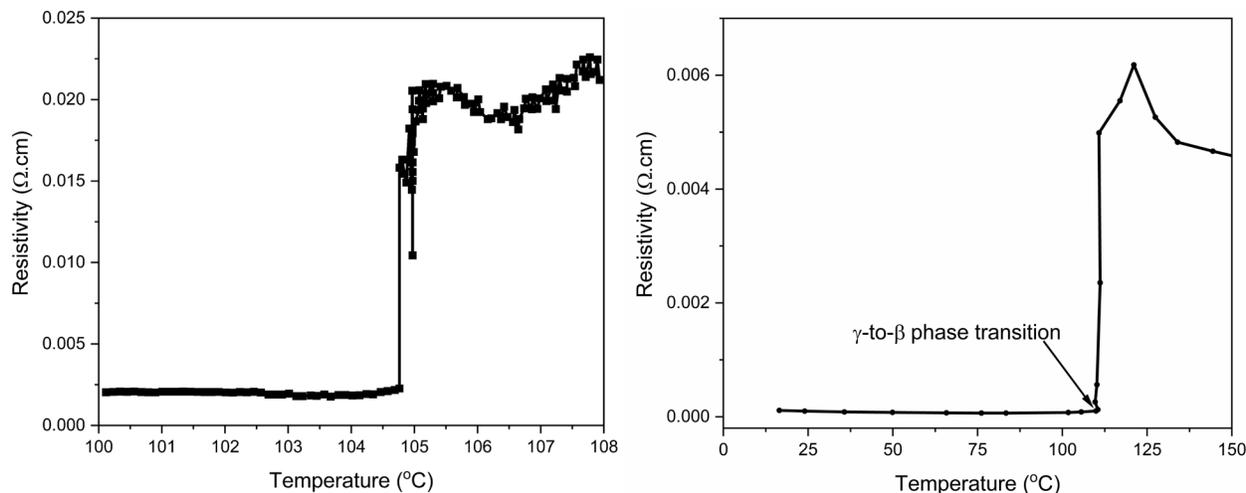

**Figure 2**. Resistivity as a function of temperature for (**left**) LK-99 digitized from the report of Lee, Kim, and co-workers,[1] and (**right**) bulk copper(I) sulfide digitized from a publication by Hirahara.[10] Note that the precise temperature of the solid–solid transition for bulk $Cu_{2-\delta}S$ deviates from 104 °C by a small magnitude depending on Cu deficiency level, $\delta$,[7,14,15] which in turn is affected by the degree of oxidation and history of electrical bias suffered by the sample. The data in the left panel are obtained from the corrected Figure 6a (inset) of ref 1 and used here under the Creative Commons Attribution Non-Commercial License. The data in the right panel are obtained from Figure 1 of ref 10 and used here with permission. Copyright 1951 The Physical Society of Japan.

The electronic conductivity, and consequently the total electrical conductivity, of $Cu_{2-\delta}S$, undergoes a sharp drop at the order–disorder transition temperature. This is because the hole mobility in the disordered β phase is significantly lower than the in the ordered γ phase. In fact, a sharp drop in electrical conductivity by two orders-of-magnitude has been reported by Hirahara.[10] Plotting this data as resistivity vs. temperature (Figure 2, right)[10] shows close correspondence with the resistivity vs. temperature plot that was reported for LK-99 as one key signature of superconductivity (Figure 2, left).[1]

As a corollary, the first-order phase transition of $Cu_{2-\delta}S$ can also lead to the temperature-dependence of the critical current ($I_c$) observed by Lee, Kim, and co-workers and interpreted[1,2] as another signature of superconductivity. The application of a constant current is likely to induce Joule heating of the sample. A sufficiently large current can heat up the sample to the temperature at which $Cu_{2-\delta}S$ undergoes the sharp transition from the high-conductivity γ phase to the low-conductivity β phase. The higher the external temperature, the lower is the current ($I_c$) that would be needed to achieve a Joule-heating-induced transition into the low-conductivity, high-resistivity β phase and observe the associated sharp rise in the voltage. The expected trend matches the observed inverse dependence of $I_c$ on the temperature (see Figure 6c in ref 1).

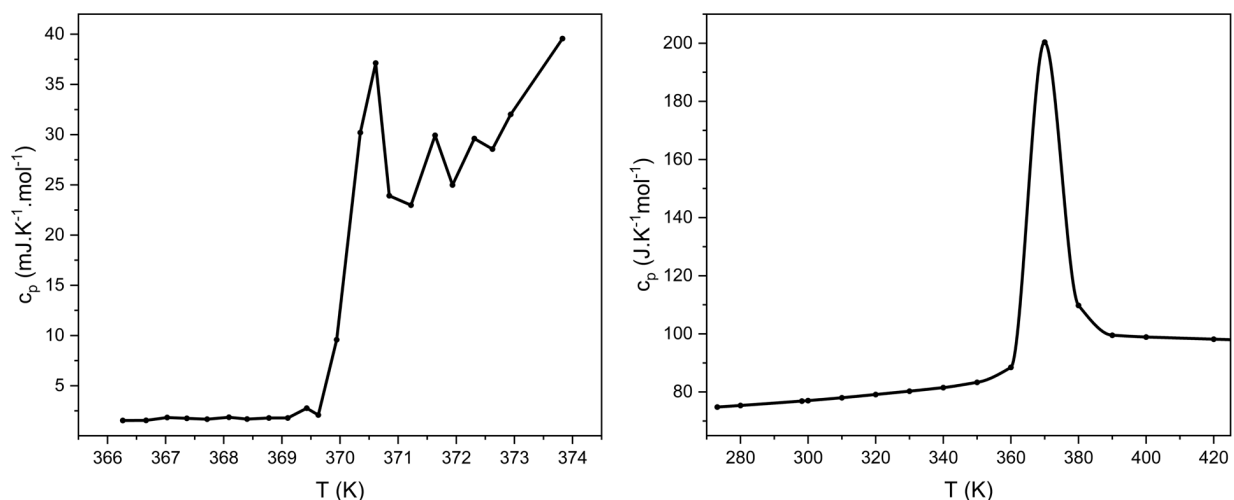

**Figure 3**. $c_p$ as a function of temperature, T, for (**left**) LK-99 digitized from the publication of Lee, Kim, and co-workers,[1] and (**right**) bulk copper(I) sulfide derived from calorimetry measurements by Gronvold et al.[16] The data in the left panel are obtained from Figure 9 of ref 1 and used here under the Creative Commons Attribution Non-Commercial License.

In their publication, Lee, Kim, and co-workers[1] also allude to a λ-transition—a hallmark of superconductivity and superfluidity—in the temperature-dependent heat capacity of LK-99. Specifically, the constant-pressure specific heat capacity, $c_p$, plotted as a function of temperature, (Figure 3, left) shows a λ-shaped feature. This feature coincides with that expected for copper(I) sulfide (Figure 3, right)[16] due to its first-order transition.

Thus, the temperature-dependent transitions in resistivity and heat capacity reported for LK-99 and identified as signatures of its superconductivity are attributable instead to the $Cu_{2-\delta}S$ byproduct present in the sample. Although one cannot rule out that the correspondence shown in Figures 2 and 3 is merely coincidental, it is clear that unambiguous investigation of the purported superconductivity of LK-99 would require testing of copper(I) sulfide-free samples. Synthesis of nearly copper(I) sulfide-free $Pb_{10-x}Cu_x(PO_4)_6O$ has just been reported and no superconducting properties were observed.[17]

The puzzle of LK-99 and its potential resolution here points to the importance of in-depth structural characterization of a new material even when such an effort appears to be of a lower priority in the face of an apparent breakthrough. This report also underscores the scientific and technological importance of obtaining and maintaining fundamental physical property data on known natural and synthetic compounds. The knowledge of the phases and phase transitions of the Cu–S system[18] that proved to be the key here was amassed decades ago starting in 1936.[19]

**Acknowledgements.** This material is based on work supported by the National Science Foundation under Grant CHE-2304910 and the author's research program on copper(I) sulfide and selenide nanostructures since 2011.

**Conflict of Interest.** The author declares no competing financial interest.